\documentclass[amsfonts,amssymb,twocolumn,showpacs,nofootinbib]{revtex4-1}

\usepackage{amsmath}
\usepackage{amsfonts}
\usepackage{amssymb}
\usepackage{mathrsfs}
\usepackage{amsthm}
\usepackage{wasysym}
\usepackage[colorlinks]{hyperref}
\usepackage[all]{hypcap}
\usepackage{graphicx}
\usepackage[usenames]{color}
\usepackage{xspace}
\usepackage{verbatim}

\usepackage[T1]{fontenc}
\usepackage{microtype}

\newcommand{\cd}{\nabla}
\newcommand{\pd}{\partial}

\newcommand{\pont}{{}^{*}\!RR}
\newcommand{\txt}[1]{{\textrm{\tiny{#1}}}}

\newcommand{\mpl}{m_\txt{pl}}

\newcommand{\eps}{\varepsilon}

\newcommand{\GR}{\txt{GR}}
\newcommand{\Def}{\txt{def}}
\newcommand{\calO}{\mathcal{O}}

\DeclareSymbolFont{euletters}{U}{eur}{m}{n}
\DeclareMathSymbol{\wp}{\mathord}{euletters}{"7D}

\begin{document}

  \title{Rapidly rotating black holes in dynamical Chern-Simons
    gravity:\newline
    Decoupling limit solutions and breakdown}
  \author{Leo C. Stein}
  \thanks{Einstein fellow}
  \email{leostein@astro.cornell.edu}
  \affiliation{Center for Radiophysics and Space Research, Cornell
    University, Ithaca, NY 14853, USA}

  \date{22 August 2014}

  \begin{abstract}
    Rapidly rotating black holes are a prime arena for understanding
    corrections to Einstein's theory of general relativity (GR).
    We construct solutions for rapidly rotating black holes in
    dynamical Chern-Simons (dCS) gravity, a useful and motivated
    example of a post-GR correction. We treat dCS as an effective
    theory and thus work in the decoupling limit, where we apply
    a perturbation scheme using the Kerr metric as the background
    solution.  Using the solutions to the scalar field and the trace
    of the metric perturbation, we determine the regime of validity of
    our perturbative approach. We find that the maximal spin limit may
    be divergent, and the decoupling limit is strongly restricted for
    rapid rotation. Rapidly-rotating stellar-mass BHs can potentially
    be used to place strong bounds on the coupling parameter $\ell$ of
    dCS.
    In order for the black hole observed in GRO~J1655-40 to be within
    the decoupling limit we need $\ell \lesssim 22$~km, a value 7
    orders of magnitude smaller than present Solar System bounds on
    dynamical Chern-Simons gravity.
  \end{abstract}


 \maketitle



\section{Introduction}
General relativity (GR), despite its
tremendous experimental and observational success~\cite{lrr-2014-4},
is widely believed to be incomplete. GR is the only classical
(non-quantum) sector of our description of nature. Semi-classical
studies suggest that the confrontation between gravity and quantum
mechanics leads to new physics~\cite{Hawking:1974rv,Almheiri:2012rt,Braunstein:2009my},
perhaps at the Planck scale.

Fundamental (top-down) approaches to a quantum resolution include string theory
and loop quantum gravity, though these approaches may be as misguided
as attempting to canonically quantize sound
waves~\cite{Jacobson:1995ab}. The bottom-up alternative is to explore
what \emph{effective} theories may arise from more fundamental ones,
and look for the phenomenology they predict. To this end there are
several well-motivated and studied corrections to GR, such as
scalar-tensor theories~\cite{Damour:1992we}, Einstein-dilaton-Gauss-Bonnet
(EdGB)~\cite{Kanti:1995vq}, and new massive gravity~\cite{deRham:2014zqa}.
In this paper we focus on dynamical Chern-Simons (dCS)
gravity~\cite{Jackiw:2003pm,Alexander:2009tp}, which is motivated from
anomaly cancellation in QFT, the low-energy limit of
string theories, or simply by being the lowest-order
parity-odd gravitational interaction. The phenomenology of dCS is
different from other low-order corrections because it is parity-odd.

Almost all corrections to GR, including dCS, involve a dimensional
coupling constant $\ell$, i.e.~an explicit length (or inverse energy)
scale.
It is this length scale $\ell$ which we want to measure or
observationally bound.
The length scale can not be too long, otherwise we might have
already noticed deviations in the weak-field (long curvature radius)
where GR is an excellent description. This motivates a perturbative
expansion in powers of the dimensionless ratio $(\ell/\mathcal{R})$,
with $\mathcal{R}$ a background length scale. This is called the
decoupling limit or weak-coupling expansion, and the leading order
equations and solutions are those of GR. In the weak-coupling regime
the theory is approximately well-posed, even if it is not
well-posed as an exact theory~\cite{Delsate:2014hba}.

\begin{figure}[b]
  \centering
  \includegraphics[width=\columnwidth{}]{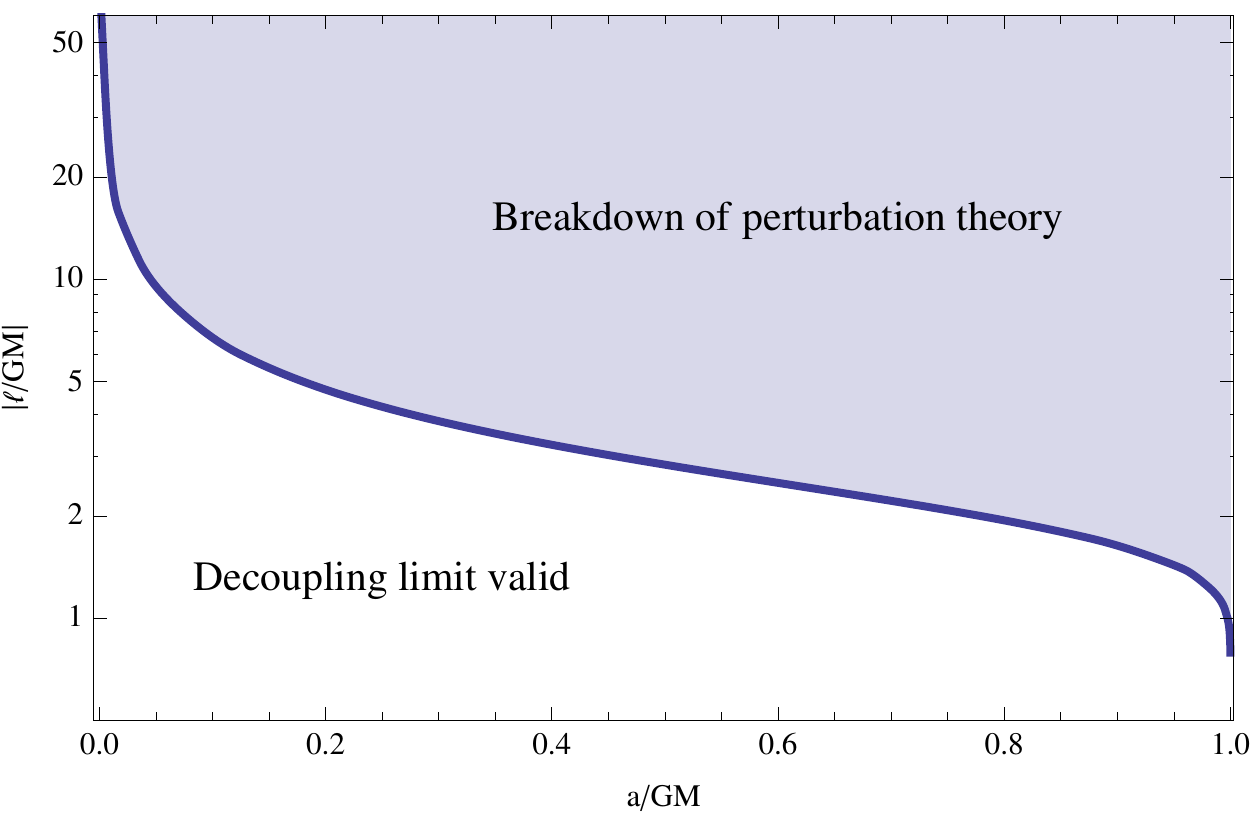}
  \caption{%
    (Color online)
    The regime of validity of the perturbation theory in the (spin,
    coupling strength) plane.  The perturbation scheme is not
    controlled in the shaded (upper) region. The small-$a$ behaviour
    can be understood analytically as $|\ell/GM|\lesssim
    \calO(a^{-1/2})$. At any $\ell$, the Chern-Simons correction
    becomes more important at higher spin, and the perturbation scheme
    eventually breaks down.
    The separatrix comes from Eq.~\eqref{eq:validity-l-htilde},
    $|\ell/GM|^{4}=1/\max |\tilde{h}(a/GM)|$.
  }
  \label{fig:validity}
\end{figure}
In this paper we study the regime of validity of the decoupling limit
in dCS by examining numerical solutions about rotating BHs, presented
in Figure~\ref{fig:validity}.
Rapidly-rotating stellar-mass BHs achieve some of the
highest-curvature gravitational fields in the present universe. As the
mass decreases and as the rotation of a BH increases, the event
horizon moves inward and the horizon curvature grows. Thus
rapidly-rotating stellar-mass BHs are a natural place to look to
understand GR or to experimentally probe for the presence of
corrections to GR. These BHs provide the strictest limits on the
applicability of the weak-coupling expansion. Present rotating BH
solutions in dCS are all in the slow-rotation
limit~\cite{Yunes:2009gi,Konno:2009kg,Pani:2011gy,Yagi:2012ya}, so
here we extend to rapid
rotation by constructing numerical solutions. This is similar to what
has been done in Ref.~\cite{Konno:2014qua}, who did not study the validity
of the weak-coupling expansion, and in Ref.~\cite{Kleihaus:2011tg} for
EdGB.

The plan for this paper is as follows.
In Sec.~\ref{sec:action-eoms-decoupling} we specify the action, lay
out the equations of motion for dCS, and develop the equations in the
decoupling limit. We pay special attention to the trace of the metric
deformation, which is simpler than the full problem, yet allows us to
determine the validity of the perturbation scheme.
In Sec.~\ref{sec:kerr-symm-reduct} we specialize to the Kerr
background geometry and rescale our functions of interest to make the
$\ell$ dependence explicit.
In Sec.~\ref{sec:num-scheme-sols} we describe our numerical scheme and
the properties of the solutions.
Finally in Sec.~\ref{sec:results-discussion} we interpret the
solutions, determining the regime of validity, and report what present
BH observations can say about $\ell$.

\section{Action, equations of motion,\\and decoupling limit}
\label{sec:action-eoms-decoupling}

We work in units where $c=1=\hbar$ so that
$[M]=[L]^{-1}$, metric signature $(-1,+1,+1,+1)$, and the sign
conventions of Wald~\cite{Wald}. We take as our action
\begin{equation}
S=\int d^{4}x\sqrt{-g}
[L_{\txt{EH}}+L_{\vartheta}+L_{\txt{int}}]
\end{equation}
with
\begin{align}
  L_{\txt{EH}} &=\frac{\mpl^{2}}{2}R\,,&
  L_{\vartheta} &=\frac{-1}{2}(\pd\vartheta)^{2} \,,&
  L_{\txt{int}} &= \frac{\mpl}{8} \ell^{2} \vartheta \ \pont\,.
\end{align}
Here $R$ is the Ricci scalar of the metric $g_{ab}$, with determinant
$g$, and the reduced Planck mass satisfies $\mpl^{2}=(8\pi G)^{-1}$.
The axionic field $\vartheta$ has been canonically normalized
and so has dimensions $[\vartheta]=[L]^{-1}$. In the interaction term
we see the Pontryagin-Chern density
\begin{equation}
  \pont = -{}^{*}\! R^{abcd} R_{abcd}
  = -\frac{1}{2}\epsilon^{abef} R_{ef}{}^{cd} R_{abcd}
\end{equation}
which is the lowest-order parity-odd curvature invariant,
constructed from the Riemann tensor $R_{abcd}$ and the
Levi-Civita tensor $\epsilon^{abcd}$.
$\pont$ is also
a topological invariant, i.e.~the integral $\int \pont \sqrt{-g}
d^{4}x$ depends only on the topology of the manifold. Finally we also
have the length scale $\ell$ which relates to the length at which this
non-minimal interaction term becomes important. $\ell$ can be thought
of as a dimensional coupling coefficient. In the limit $\ell\to 0$,
general relativity is recovered. It is this length scale $\ell$ which
in principle could be observationally constrained.

Variation of this action with respect to $\vartheta$ leads to the
scalar equation of motion
\begin{equation}
  \label{eq:eom-vartheta}
  \square\vartheta = - \frac{\mpl}{8} \ell^{2} \ \pont\,,
\end{equation}
where $\square=\cd^{a}\cd_{a}$ and $\cd_{a}$ is the covariant
derivative compatible with the metric.
Variation of the action with respect to the inverse metric
$g^{ab}$ leads to the metric equation of motion
\begin{equation}
  \label{eq:eom-metric}
  \mpl^2 G_{ab} +\mpl\ell^{2}  C_{ab} =
  T^{(m)}_{ab}+T^{(\vartheta)}_{ab}\,.
\end{equation}
Here we have the stress-energy tensor of any matter fields,
$T^{(m)}_{ab}$, and the stress-energy tensor of the canonical scalar
with flat potential,
\begin{equation}
T^{(\vartheta)}_{ab} = \cd_a \vartheta \cd_b \vartheta -
{\textstyle\frac{1}{2}}g_{ab}\cd^c\vartheta\cd_c\vartheta  \,.
\end{equation}
We also have the $C$-tensor, with the convention~\cite{Alexander:2009tp}
(hence the factor of $\frac{1}{8}$ in the action)
\begin{equation}
  \label{eq:C-def}
  C_{ab} = \epsilon^{cde}{}_{(a}R_{b)c;d}\vartheta_{;e}
  +{}^*\!R^{c}{}_{(a}{}_{b)}{}^{d}\vartheta_{;cd}\,.
\end{equation}
The $C$-tensor is trace-free, $g^{ab}C_{ab}=0$, and it satisfies the
divergence identity
\begin{equation}
\label{eq:div-C}
\cd^a \left(T^{(\vartheta)}_{ab} - \mpl\ell^{2} C_{ab}\right) =
\left(\square\vartheta + \frac{\mpl}{8}\ell^{2}\pont\right)
\cd_b\vartheta = 0\,.
\end{equation}
Thus we have ordinary conservation of matter stress-energy,
$\cd^{a}T^{(m)}_{ab}=0$. In this paper we will not consider any matter
sources, $T_{ab}^{(m)}=0$.

We now take the decoupling limit, where we assume that the corrections
due to the interaction term are ``small.'' This allows us to perform a
controlled, perturbative expansion of all the fields in terms of the
coupling strength $\ell$. We will introduce a formal order-counting
parameter $\eps$ to keep track of the perturbation scheme, which can
be set to 1 later, which counts the order in $\ell^{2}$. That is, we take
$\ell^{2}\to\eps\ell^{2}$ and expand both the metric and scalar in
powers of $\eps$: $\vartheta = \sum_{k=0}^{\infty} \eps^{k}
\vartheta^{(k)}/k!$ and similarly for the metric. In order to recover
the GR solution in the limit $\eps\to0$, we have that
$\vartheta^{(0)}=0$, and $g_{ab}^{(0)}=g_{ab}^{\GR}$ for some
known solution.

From Eq.~\eqref{eq:eom-vartheta}, we can see that the leading order
solution for $\vartheta$ is $\vartheta^{(1)}$, which satisfies
\begin{equation}
  \label{eq:eom-vartheta-1}
  \square^{(0)}\vartheta^{(1)} = - \frac{\mpl}{8} \ell^{2} \ \pont^{(0)}\,.
\end{equation}
From here forward we will drop the superscript $(0)$ when it is
unambiguous.
Now analyze
Eq.~\eqref{eq:eom-metric} with $\ell^{2}\to\eps\ell^{2}$,
and recall that $C_{ab}$ is linear in
$\vartheta$, while $T_{ab}^{(\vartheta)}$ is quadratic in
$\vartheta$. This shows that $h_{ab}^{(1)}$ has vanishing source
term, and the leading order metric deformation away from GR enters at
$\eps^{2}$:
$g_{ab}=g_{ab}^{\txt{GR}}+\eps^{2}h_{ab}^{(2)}/2+\calO(\eps^{3})$.
We label this as $h_{ab}^{(2)}/2\equiv
h_{ab}^{\Def}$, which satisfies
\begin{equation}
  \label{eq:eom-h-Def}
  \mpl^2 G_{ab}^{(1)}[ h_{cd}^{\Def}] +\mpl\ell^{2}  C_{ab}[\vartheta^{(1)}] =
  T^{(\vartheta)}_{ab}[\vartheta^{(1)},\vartheta^{(1)}]\,,
\end{equation}
where $G_{ab}^{(1)}[h_{cd}]$ is the linearized Einstein operator acting on
the metric perturbation $h_{cd}$.

Our main concern in this paper is the regime of validity of the
decoupling limit
via this perturbation scheme. To quantify if this scheme is under
control, we must check that some appropriate dimensionless quantities
are ``small.'' For example, in~\cite{Stein:2013wza} it was possible to
make scaling arguments for the ratio of the interaction Lagrangian
$L_{\txt{int}}$ and the Einstein-Hilbert Lagrangian $L_{\txt{EH}}$
when in the presence of a matter source. However, this is not possible
here because we are dealing with a matter-free background solution, so
the E-H Lagrangian identically vanishes. Similarly, we can not make a
comparison of $T^{(\vartheta)}_{ab}$ to $G_{ab}$, because matter-free
backgrounds give a vanishing Einstein tensor.
Instead, we investigate what is possible
with the metric deformation $h_{ab}^{\Def}$.

Clearly it is much harder to solve the metric deformation
equation~\eqref{eq:eom-h-Def} than the scalar equation
Eq.~\eqref{eq:eom-vartheta-1}. We will not attempt to solve for a full
metric tensor deformation solution. Rather, note what is possible when
tracing (with $g^{ab}_{\GR}$) the metric deformation equation. With a
Ricci-flat background and in the Lorenz gauge
($\cd^{b}h_{ab}^{\Def}=\frac{1}{2}\cd_{a}h^{\Def}$) we find
\begin{equation}
  \label{eq:eom-trace}
  \frac{1}{2}\mpl^{2}\square h^{\Def}
  = - (\cd^{a}\vartheta^{(1)})(\cd_{a}\vartheta^{(1)})\,,
\end{equation}
where $h^{\Def}=g^{ab}_{\GR}h_{ab}^{\Def}$ is the trace of the metric
perturbation. This is just another sourced scalar d'Alembertian equation, the same
type of equation we must solve for the the scalar field
$\vartheta^{(1)}$.

Once we find a solution for $h^{\Def}$, we still have to
find an appropriate dimensionless comparison in order to verify that
the perturbation scheme is under control. Consider the perturbative
expansion of the volume element:
\begin{equation}
  \sqrt{-g} = \sqrt{-g^{\GR}}(1+ \eps^{2}{\textstyle\frac{1}{2}}h^{\Def})+\calO(\eps^{3})\,.
\end{equation}
If the quantity $h^{\Def}$ becomes $\calO(1)$, then
clearly we should keep higher order terms in this
expansion.
Although this statement
is gauge-dependent, it gives an order of magnitude estimate of
the regime of validity of the perturbation scheme. There are also
gauge-invariant quantities that can be constructed from $h^{\Def}$,
such as the perturbed 4-volume of a region. Moreover, the
BH-spin-dependent structure of the regime of validity will still be
revealed with this condition. Therefore we define our criterion for
the perturbation to be under control:
\begin{equation}
  \label{eq:validity-condition}
|h^{\Def}|\lesssim 1\,.
\end{equation}

\section{Kerr, symmetry reduction,\\and scaling}
\label{sec:kerr-symm-reduct}

We are seeking rapidly rotating black hole solutions, and we have the
luxury that at $\calO(\eps^{0})$ our solution reduces to the one in
GR. Therefore we take the Kerr metric, $g_{ab}^{\GR}=g_{ab}^{K}$, which in
Boyer-Lindquist coordinates is~\cite{Wald}
\begin{align}
  g^{K}_{ab} dx^{a} dx^{b} =&
-\frac{\Delta}{\Sigma}\left(dt - a \sin^2\theta d\phi \right)^2 +
\Sigma \left(\frac{dr^2}{\Delta} + d\theta^2\right) \nonumber\\
&+\frac{\sin^2\theta}{\Sigma}\Big((r^2+a^2)d\phi - a dt\Big)^2 \,,
\end{align}
with the total mass $M$, angular momentum per unit mass $a=J/M$ (with
units $[a]=[L]$), $-GM<a<+GM$, and where
\begin{align}
  \label{eq:sigma}
  \Sigma &\equiv r^2+a^2\cos^2\theta \\
  \label{eq:Delta}
  \Delta &\equiv r^2+a^2-2GMr\,.
\end{align}
The event horizon is at the outer root $r_{+}$ of $\Delta=0$, given by
$r_{\pm}=GM\pm\sqrt{(GM)^{2}-a^{2}}$. In these coordinates the root of
the metric determinant is given by
\begin{equation}
  \sqrt{-g^{K}} = \Sigma\sin\theta\,.
\end{equation}
A straightforward calculation gives
\begin{equation}
  \pont = 96 (GM)^{2}\frac{a\mu r(3r^{2}-a^{2}\mu^{2})(r^{2}-3a^{2}\mu^{2})}{\Sigma^{6}} \,,
\end{equation}
where we use the shorthand $\mu=\cos\theta$.

The d'Alembertian operator $\square$ in the Kerr background
is somewhat complicated, but since we are seeking stationary and
axisymmetric solutions, the operator simplifies:
\begin{align}
  \square f(r,\theta) &= \frac{1}{\sqrt{-g}}\pd_{a}
  \left(
\sqrt{-g} g^{ab} \pd_{b} f(r,\theta)
  \right) \,,\\
 \label{eq:sigma-box-f-kerr}
  \Sigma \square f(r,\theta) &=
  \left[\pd_{r} \Delta \pd_{r} +
  \pd_{\mu}(1-\mu^{2})\pd_{\mu}
  \right] f\,,
\end{align}
where $\pd_{\mu} = \frac{\pd}{\pd\cos\theta} =
\frac{-1}{\sin\theta}\frac{\pd}{\pd\theta}$.

Before discussing how to solve this partial differential equation
(PDE), let us rescale to
dimensionless coordinates, which will also elucidate the mass
dependence of the solution. Let $\tilde{r}\equiv r/GM$,
$\tilde{a}\equiv a/GM$, $\tilde{\Delta}\equiv\Delta/(GM)^{2} =
\tilde{r}^{2}+\tilde{a}^{2}-2\tilde{r}$,
$\tilde{\Sigma}\equiv\Sigma/(GM)^{2} =
\tilde{r}^{2}+\tilde{a}^{2}\mu^{2}$.  Analyzing the equation of motion
for $\vartheta^{(1)}$ we find that we should rescale it as
\begin{equation}
  \label{eq:theta-tilde-def}
  \vartheta^{(1)} = \tilde{\vartheta} \mpl \frac{\ell^{2}}{(GM)^{2}}\,.
\end{equation}
Then the equation for $\tilde{\vartheta}$
is
\begin{multline}
  \label{eq:eom-thetatilde}
  \left[
    \pd_{\tilde{r}} \tilde{\Delta} \pd_{\tilde{r}} +
    \pd_{\mu}(1-\mu^{2})\pd_{\mu}
  \right]\tilde{\vartheta} =\\
- 12\frac{\tilde{a}\mu \tilde{r}
    (3\tilde{r}^{2}-\tilde{a}^{2}\mu^{2})
    (\tilde{r}^{2}-3\tilde{a}^{2}\mu^{2})}{\tilde{\Sigma}^{5}} \,.
\end{multline}
From the scaling in Eq.~\eqref{eq:theta-tilde-def} we see some obvious
features. As $\ell$ increases, $\vartheta^{(1)}$ increases, which is easy to
understand since $\ell$ is acting as a coupling strength. We also see
that black holes with lighter masses generate larger values of
$\vartheta^{(1)}$. That is, the CS interaction is a
higher-curvature operator, which is important at shorter lengths,
which corresponds to lighter black holes. In other words, the
curvature at the horizon of a black hole goes as $1/M^{2}$, so the
greatest effect comes from the lightest black hole.

Now we perform another scaling analysis in the equation for
$h^{\Def}$, Eq.~\eqref{eq:eom-trace}. What we find is that we should
introduce the scaling
\begin{equation}
  h^{\Def} = \tilde{h}
  \left(\frac{\ell}{GM}\right)^{4}\,,
\end{equation}
where $\tilde{h}$ satisfies
\begin{multline}
  \label{eq:eom-htilde}
    \left[
    \pd_{\tilde{r}} \tilde{\Delta} \pd_{\tilde{r}} +
    \pd_{\mu}(1-\mu^{2})\pd_{\mu}
  \right]\tilde{h} =\\
  -2
  \left[
    \tilde{\Delta} (\pd_{\tilde{r}}\tilde{\vartheta})^{2} + (1-\mu^{2}) (\pd_{\mu}\tilde{\vartheta})^{2}
  \right]\,.
\end{multline}
We again see that as $\ell$ increases and/or $M$ decreases, $h^{\Def}$
increases. In terms of the scaled variable $\tilde{h}$, the validity
condition [Eq.~\eqref{eq:validity-condition}] reads
\begin{equation}
  \left|
    \tilde{h} \left(\frac{\ell}{GM}\right)^{4}
  \right|\lesssim 1\,.
\end{equation}
Alternatively, if a black hole with some known spin $\tilde{a}$ is found to be
well-described everywhere by GR, we may claim the condition
\begin{equation}
  \label{eq:validity-l-htilde}
  \left|
    \frac{\ell}{GM}
  \right|^{4}\lesssim \frac{1}{\max |\tilde{h}(\tilde{a})|}\,.
\end{equation}
From this we see that lighter black holes will produce better
bounds on $\ell$, as will more rapidly-rotating black holes.
This latter point comes from noting that in the small
$\tilde{a}$ expansion, $\tilde{h}\sim\calO(\tilde{a}^{2})$; thus more
rapidly spinning black holes source a larger metric deformation and
provide a more stringent bound on $\ell$. Our goal now is to determine
$\max|\tilde{h}(\tilde{a})|$ by solving the system comprising
Eqs.~\eqref{eq:eom-thetatilde} and \eqref{eq:eom-htilde}.

\section{Numerical scheme and solutions}
\label{sec:num-scheme-sols}

Several approaches are available to try to solve the system of
equations. Each equation is an elliptic PDE on the exterior domain
$\tilde{r}\in[\tilde{r}_{+},+\infty)$, $\mu\in[-1,+1]$. The factor of
$\tilde\Delta$ in the $\pd_{\tilde{r}}^{2}$ goes to zero as
$\tilde{r}\to\tilde{r}_{+}$, and the equation changes from an elliptic
to a hyperbolic one inside the (outer) event horizon.

The first approach to solving this system was made
in~\cite{Campbell:1990ai} for $\vartheta^{(1)}$
and~\cite{Yunes:2009gi,Konno:2009kg} for $h_{ab}^{\Def}$, by expanding all fields in a
bivariate
expansion in $\eps$ and $a$ and finding the leading (linear in $a$)
solution. This was further extended in~\cite{Yagi:2012ya} to quadratic
order in $a$. However, we are interested in the full behaviour in $a$,
not just a slow-rotation expansion. There is no guarantee that an
expansion about $a=0$ will converge as $\tilde{a}\to 1$. In fact, our
numerical results will suggest that the $\tilde{a}\to 1$ limit is singular,
which restricts the radius of convergence of an $a$ expansion.

The wave equation on the Kerr background is amenable to separation of
variables, as was demonstrated in the celebrated work of
Teukolsky~\cite{Teukolsky:1972my,Teukolsky:1973ha,1974PhDT.......103T}. That
feature is naturally retained here, and in terms of the separation
approach, these solutions would have support only for vanishing
temporal Fourier frequency $\omega=0$
(stationary) and azimuthal number $m=0$ (axisymmetric). This approach has recently been
attempted by Konno and Takahashi~\cite{Konno:2014qua}. In the
separation of variables approach, the angular basis functions are
simply Legendre polynomials $P_{j}(\mu)$; both the source terms and
solutions are expanded in this basis.
It is straightforward to see that the solution for $\vartheta^{(1)}$
will have support only at odd $j$, while the solution for $h^{\Def}$
has support only at even $j$. Konno and Takahashi gave formulae for
the moments of the source $\Sigma \pont$ as a quadruple sum by
expanding everything in powers of $\mu$.
Using a different approach, this author has presented a more compact
form~\cite{Stein:2014wza} in terms of rational polynomials of $r$
times hypergeometric functions. These moments
would then have to be integrated against the homogeneous solutions to
the radial ODE for each mode. These homogeneous solutions are Legendre
functions of the first and second kind, respectively $P_{j}(\eta)$ and
$Q_{j}(\eta)$, where
\begin{align}
  \label{eq:eta-def}
  \eta&\equiv (r-GM)/b\,, &  b&=\sqrt{(GM)^{2}-a^{2}}\,,
\end{align}
with $\eta\in[+1,+\infty)$.
However, neither Konno and Takahashi nor the present author have found
general expressions for the radial indefinite integrals.

Yet another analytic approach is to try to integrate the source term
against the analytically known Green's
function~\cite{Ottewill:2012aj}. However, the Green's function is
written in terms of a complete elliptic integral of the first
kind~\cite{TaylorsEmail}, and the present author has not had any success
integrating the source against the Green's function. We are unaware of
any such attempt in the literature.

Instead, we solve the system of equations numerically. Still there are
several approaches available. One may use a hyperbolic (wave equation)
solver with some initial guess and just wait for the transient
solution to settle down to the stationary one. Alternatively, one may
use a relaxation scheme to solve the purely elliptic problem. In
between these two approaches, one may add a $-\pd_{t}\tilde\vartheta$
term to the elliptic operator to make it parabolic, thus approximating
an exponentially convergent relaxation scheme. All of these approaches
are workable.

We will take advantage of the separability of our equation in order to
directly invert the differential operator. This is the numerical
analog of the Green's function approach, but applied mode-by-mode in
the spectral decomposition.

Consider the same differential operator with any given source term,
\begin{equation}
  \left[
    \pd_{\tilde{r}} \tilde{\Delta} \pd_{\tilde{r}} +
    \pd_{\mu}(1-\mu^{2})\pd_{\mu}
  \right] f = S\,.
\end{equation}
Now expand both the source term $S$ and the solution we seek $f$ in
terms of Legendre polynomials,
\begin{align}
  S(\tilde{r},\mu) &= \sum_{j=0}^{\infty} S_{j}(\tilde{r}) P_{j}(\mu)\,, 
\end{align}
where
\begin{align}
  S_{j}(\tilde{r}) = \frac{2j+1}{2}\int_{-1}^{+1} S(\tilde{r},\mu)
  P_{j}(\mu) d\mu
\end{align}
and similarly for $f$, with the prefactor arising from the normalization
\begin{equation}
\int_{-1}^{+1} P_{j}(\mu) P_{j'}(\mu) d\mu = \frac{2}{2j+1}\delta_{jj'} \,.
\end{equation}
This defines the forward and backward spectral transformation for the
angular direction of the domain.

Now each $f_{j}$ satisfies an ODE,
\begin{equation}
  \left[
\pd_{\tilde{r}} \tilde{\Delta} \pd_{\tilde{r}}  - j(j+1)
  \right] f_{j}(\tilde{r}) = S_{j}(\tilde{r})\,.
\end{equation}
We use a pseudospectral collocation scheme~\cite{boyd2001chebyshev} to
directly invert the differential operator appearing here. First, we remap
the radial domain, using the definition of the dimensionless
$\eta$ [Eq.~\eqref{eq:eta-def}], so that the radial domain is
$a$-independent. Then we compactify via
\begin{equation}
  \label{eq:x-eta-mapping}
  \eta = \frac{2}{1-x}\,,
\end{equation}
so that $x\in[-1,+1]$ with the horizon at $x=-1$ and spatial infinity
at $x=+1$. In terms of this new radial coordinate, we have
\begin{equation}
  \label{eq:ode-in-x}
  \left[
{\textstyle \frac{1}{4}}(1-x)^{2} \pd_{x} (3-x)(1+x)\pd_{x}
-j(j+1)
  \right] f_{j}(x) = S_{j}(x)\,.
\end{equation}
Besides being compact, this coordinate has another advantage: the
radial operator which we want to invert is now $a$-independent, which
means the same solver [e.g.~a lower-upper (LU) decomposition] can be precomputed
once and applied for all $a$. Now we use Chebyshev polynomials as
basis functions and the interior (``roots'') grid for the collocation
points for our pseudospectral method~\cite{boyd2001chebyshev} when
solving Eq.~\eqref{eq:ode-in-x}.

The physical boundary conditions are regularity at both the horizon
and infinity, which means that $\tilde\vartheta$ and $\tilde{h}$ must
vanish at $x=+1$. Recall that the homogeneous solutions (which
comprise the null space of the full differential operator) are
$P_{j}(\eta)$ and $Q_{j}(\eta)$. Since these blow up at one endpoint,
they are automatically excluded by the discretized scheme (i.e.~the
discretized differential operator is full rank), except for
$P_{0}(\eta)$. Therefore we must include one numerical boundary
condition for the $j=0$ mode to enforce vanishing at $x=+1$.

For the highest spins we have investigated, $\tilde{a}=0.99995$, we
have found that 55 angular basis functions and 64 radial grid points
are sufficient to recover $\tilde{h}(x,\mu)$ with fractional errors
estimated at the relative $3\times 10^{-8}$ level. Meanwhile, this
resolution recovers solutions at lower spins (below $\tilde{a}\lesssim
0.998$) at the relative $3\times 10^{-12}$ level. Thus each $j$ mode
requires solving a linear system whose dimension is just 64. This
scheme is implemented in \textsc{Mathematica}, and solves for
$\tilde{\vartheta}$ and $\tilde{h}$ for a single value of $\tilde{a}$ in
$0.19$~sec on a laptop computer.

\begin{figure}[b]
  \centering
  \includegraphics[width=\columnwidth]{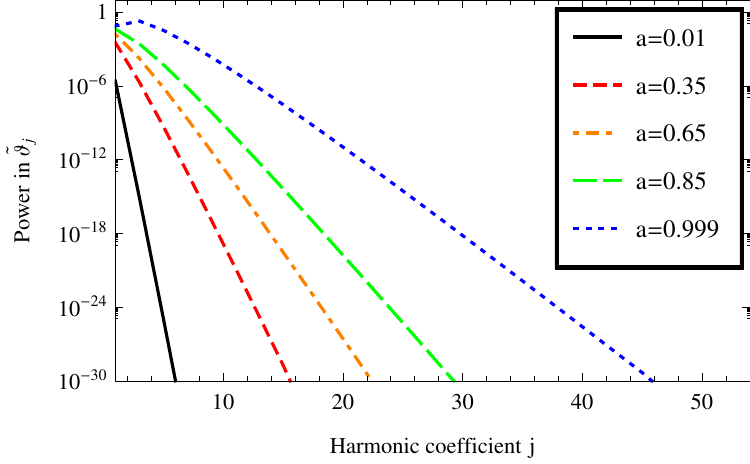}
  \caption{
    (Color online)
    Smooth solutions for $\tilde\vartheta$ and $\tilde h$ have
    exponential convergence as a decomposition in Legendre
    polynomials $P_{j}(\cos\theta)$.
    The vertical axis is the $L^{2}$ norm of
    $\tilde{\vartheta}_{j}(x)$ [Eq.~\eqref{eq:L2-norm}].
    At low spin, the convergence is more rapid, and few coefficients
    are needed. As spin increases more coefficients must be kept.
    Only the odd coefficients of $\tilde\vartheta$ are plotted.
    $\tilde{h}$ follows the same trend.
  }
  \label{fig:power-theta}
\end{figure}

The distribution of power into the $P_{j}$ modes, as a function of the
spin $\tilde{a}$, is seen in Fig.~\ref{fig:power-theta}. For each
solution $\tilde{\vartheta}_{j}(x)$ at a given spin, we plot a
discrete estimate of the $L^{2}$ norm,
\begin{equation}
  \label{eq:L2-norm}
  \left\|\tilde{\vartheta}_{j}(x)\right\|_{2} = \int_{-1}^{+1}
  |\tilde{\vartheta}_{j}(x)|^{2} dx \,.
\end{equation}
For the odd function $\tilde\vartheta$, we only plot the power in odd
$j$ coefficients. The plot for the even function $\tilde{h}$ looks
similar, when keeping just the even $j$'s. As expected, there is
exponential convergence in $P_{j}$ coefficients, because the sources
and solutions are $C^{\infty}$. However, there is a striking feature
in the spin-dependence of this exponential convergence. As spin
increases, the rate of convergence decreases: more $P_{j}$
coefficients are needed for the same accuracy. Finally, at
sufficiently high $\tilde{a}$, the peak power is no longer in
$\tilde\vartheta_{1}$, as can be seen in the line for
$\tilde{a}=0.999$.

\begin{figure*}[tbp]
  \centering
  \includegraphics[width=\textwidth]{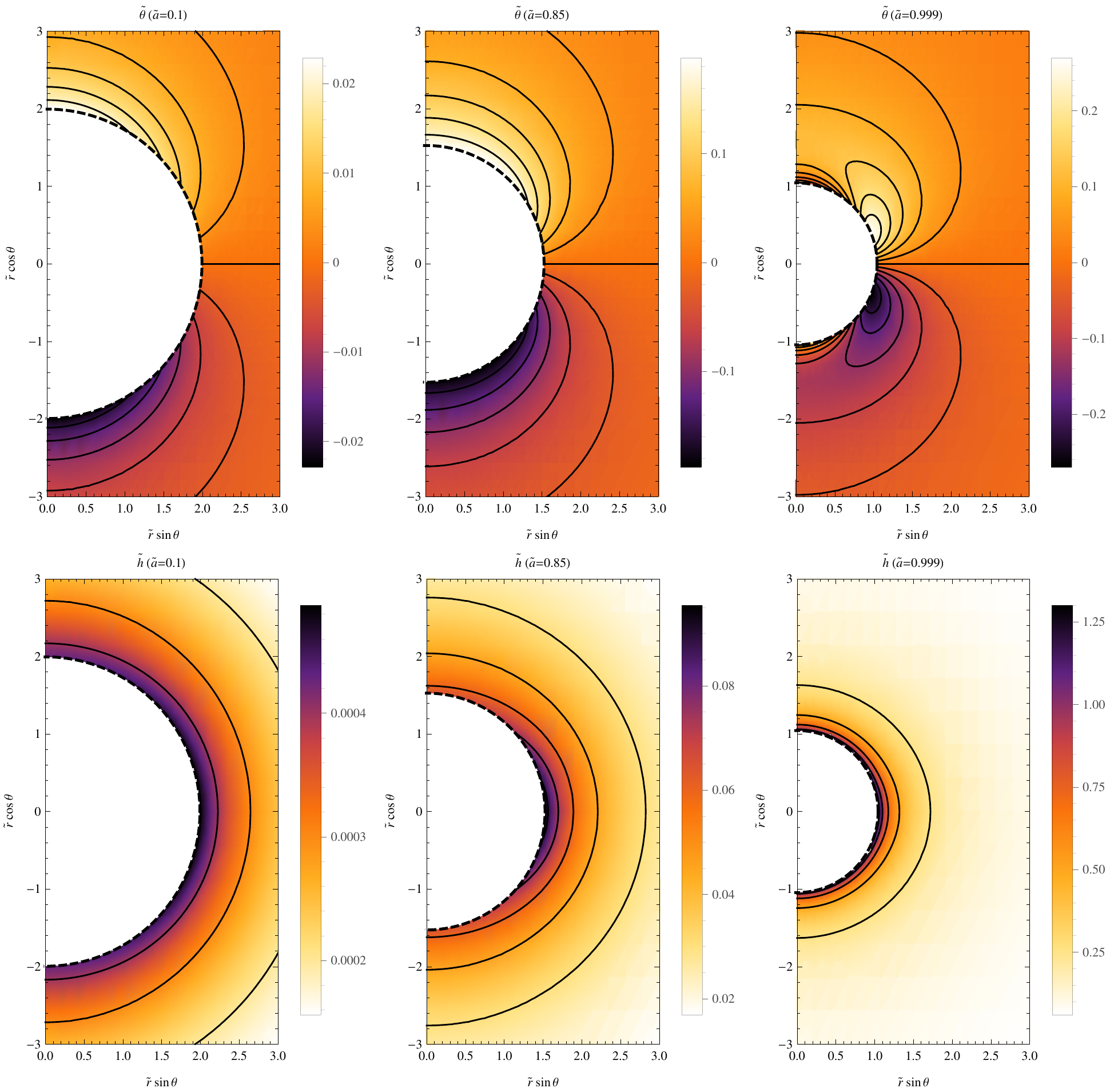}
  \caption{
    (Color online)
    Profiles of solutions for $\tilde{\vartheta}$ (top panels) and
    $\tilde{h}$ (bottom panels) in a longitudinal ($\phi=\text{const.}$)
    section of the space. From left to right, the profiles are at low
    spin ($\tilde{a}=0.1$), intermediate ($\tilde{a}=0.85$),
    and high spin ($\tilde{a}=0.999$). Color represents the value of
    the field. Note the different color bar scale for each panel.
    Contours of constant field value are spaced linearly.
    The dashed line represents the horizon.
    At low spin, the $\tilde{\vartheta}$ solution is almost a pure
    dipole solution, $\propto P_{1}(\cos\theta)$. At intermediate and
    higher spin, the solutions develop more multipole structure.
    $\tilde{h}$ is always highly peaked on the horizon at the equator,
    $\cos\theta=0$. This is seen more easily in
    Fig.~\ref{fig:h-3d-plot}.
  }
  \label{fig:profilesGrid}
\end{figure*}

Next we look at the spatial structure of some of the solutions in
Fig.~\ref{fig:profilesGrid}. The solutions are azimuthally symmetric,
so we plot only a longitudinal section, with color corresponding to
the value of the field (notice that the color scale differs for each
panel). In the $(\tilde{r}\sin\theta, \tilde{r}\cos\theta)$
coordinates we use, the horizon is simply a circle (these are not the
quasi-Cartesian Kerr-Schild coordinates, so they do not show the
oblateness of the spacetime). The contours of constant field value
(linearly spaced) help to highlight the multipole structure. The high
multipole content is easily seen in the top-right panel.

\begin{figure}[tb]
  \centering
  \includegraphics[width=\columnwidth]{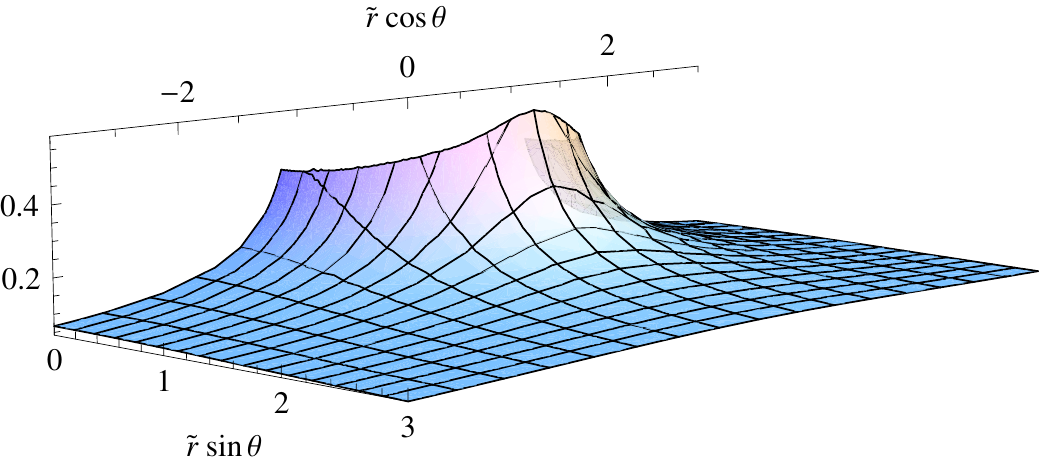}
  \caption{
    (Color online)
    Surface representing the solution for $\tilde{h}$ at
    $\tilde{a}=0.99$ on a longitudinal ($\phi=\text{const.}$) section
    of the space. The two horizontal directions are
    $\tilde{r}\cos\theta$ and $\tilde{r}\sin\theta$ within the
    section, and the vertical height represents the value of
    $\tilde{h}(\tilde{r},\theta)$.  The solution peaks very strongly
    near the horizon. For all values of $\tilde{a}$, the maximum of
    $\tilde{h}$ occurs on the equator at the horizon. This is the
    value which enters into Eq.~\eqref{eq:validity-l-htilde} and thus
    the separatrix on Fig.~\ref{fig:validity}.
  }
  \label{fig:h-3d-plot}
\end{figure}

We care most about the spin-dependence of
$\max|\tilde{h}(\tilde{a})|$. As we can see in the bottom row of
Fig.~\ref{fig:profilesGrid}, $\tilde{h}$ always attains its peak value
on the horizon, at the equator ($\cos\theta=0$). Because of the very
strong radial dependence of the solution, this is difficult to see in
the color density plots. It is made more apparent in the 3D surface
plot of Fig.~\ref{fig:h-3d-plot}. There we use the same
$(\tilde{r}\sin\theta, \tilde{r}\cos\theta)$ coordinates for the
horizontal directions, and turn the field value of $\tilde{h}$ into
the vertical height of the surface. We have plotted the solution for
$\tilde{a}=0.99$, but the qualitative structure is similar for other
spins: the peak value is at the horizon, on the equator.

\section{Results and discussion}
\label{sec:results-discussion}

We now turn to determining the regime of validity of the perturbation
scheme. From Eq.~\eqref{eq:validity-l-htilde}, we see that we need
to determine $\max|\tilde{h}(\tilde{a})|$. From the numerical scheme
described above, we see that $\tilde{h}$ attains its maximum value on
the horizon, at the equator. It is straightforward to extract this
value from the numerical solutions. For this purpose we ran 29 models
at different values of $\tilde{a}$, which takes about $5.5$~sec on a
laptop computer. We concentrated more models towards the
endpoints ($\tilde{a}=0$ and 1) where $\max|\tilde{h}(\tilde{a})|$
changes most rapidly with $\tilde{a}$. It is then simple to convert
this to the separatrix between the regime of validity and breakdown
through $|\ell/GM|^{4} = 1/\max |\tilde{h}(a/GM)|$.

These results are presented in Fig.~\ref{fig:validity}. As expected,
larger values of $\tilde{a}$ induce a larger Chern-Simons modification, and
thus the range of $\ell/GM$ where the perturbation scheme is valid is
smaller. The small-$\tilde{a}$ behaviour can be easily understood
analytically. Recall that for small spin, $\tilde{\vartheta}\propto
\tilde{a}$, and then $\tilde{h}\propto\tilde{a}^{2}$. Taking the
$-1/4$ power to convert to the $\ell$-separatrix, we have that
$|\ell/GM|_{\txt{sep}} \propto \tilde{a}^{-1/2}$ for small
$\tilde{a}$.

There is also clearly a feature at $\tilde{a}\to 1$. We do not have an
analytic explanation for this feature, since analytic results are only
available up to $\calO(\tilde{a}^{2})$ as a power series
expansion, and we have no analytic expansion about $\tilde{a}\to 1$. Close
examination of the maximum spin limit suggests that
this limit may in fact be divergent. One possible explanation
is as follows: Konno and Takahashi have suggested~\cite{Konno:2014qua}
that the scalar field solution blows up at the inner horizon. Recall
that as $\tilde{a}\to 1$, the inner horizon approaches the outer
horizon. This may lead to a divergence in $\tilde{\vartheta}$, and
a worse divergence in $\tilde{h}$, since the source for $\tilde{h}$ is
constructed from $(\pd\tilde{\vartheta})^{2}$.


This strong spin dependence at the high-spin end has important
implications for observationally constraining the coupling length
$\ell$. There are now several black hole systems known with spins
approaching maximal values. One very promising candidate is GRO~J1655-40,
with a mass $M=6.30\pm0.27 M_{\odot}$ and a spin
$\tilde{a} \approx 0.65$--$0.75$~\cite{Shafee:2005ef}. If it were
observationally possible to infer that this system is well described everywhere
by general relativity (i.e.,~that deviations from GR are small), then
we could make the claim that
\begin{equation}
  \ell \lesssim 15~GM_{\odot} \approx 22~\text{km}\,.
\end{equation}
Such a bound would improve on present Solar System
constraints~\cite{AliHaimoud:2011fw} by seven orders of
magnitude, and is comparable in magnitude to that forecasted
by~\cite{Yagi:2012vf}.
We must caution the reader, though, that such bounds will be highly
dependent on modeling of the physics of the accretion disk, which is
uncertain.

\subsection*{Future work}

As we have mentioned above, there are still several outstanding issues
which warrant further investigation. We do not have an analytic
explanation of the $\tilde{a}\to 1$ limit seen in
Fig.~\ref{fig:validity}. Just as it's possible to perform a slow-spin
expansion, it should be possible to perform a near-extremal expansion
to get a better understanding of this limit.
Perhaps it is possible to obtain a fully analytic solution valid
for all $a$.

The bounds we have forecasted here are just estimates, and real bounds
would require much more work. Most important is an understanding of
the impact of the dCS correction on the electromagnetic observables
from BH binary systems, and how degenerate is the signature with
modeling uncertainty of the accretion physics.
This certainly requires a full metric solution, not just the trace.

Stepping back further, there is another regime of validity which we
should discuss. Here we have concerned ourselves with the validity of
the perturbation scheme for the decoupling limit. There is also the
validity or breakdown of the effective field theory: the action is
thought of as an $\ell$ expansion of some higher-energy theory, and we
have truncated at some order. The two regimes of validity are not the
same. A na\"ive estimate of the EFT breakdown is at $|\ell|\sim GM$,
which does not involve spin at all. Note that at sufficiently high
spin, the weak-coupling breakdown crosses the EFT breakdown.
We hope to investigate how the EFT breakdown relates to our results.

\acknowledgments

The author acknowledges Barry Wardell, \'Eanna Flanagan,
David Nichols, Peter Taylor, Kent Yagi, and Nico Yunes
for useful discussion.
LCS acknowledges that support for this work was
provided by the National Aeronautics and Space Administration through
Einstein Postdoctoral Fellowship Award Number PF2-130101 issued by the
Chandra X-ray Observatory Center, which is operated by the Smithsonian
Astrophysical Observatory for and on behalf of the National
Aeronautics Space Administration under contract NAS8-03060.




\bibliographystyle{apsrev4-1}
\bibliography{CS-Kerr-scalar}

\begin{thebibliography}{30}%
\makeatletter
\providecommand \@ifxundefined [1]{%
 \@ifx{#1\undefined}
}%
\providecommand \@ifnum [1]{%
 \ifnum #1\expandafter \@firstoftwo
 \else \expandafter \@secondoftwo
 \fi
}%
\providecommand \@ifx [1]{%
 \ifx #1\expandafter \@firstoftwo
 \else \expandafter \@secondoftwo
 \fi
}%
\providecommand \natexlab [1]{#1}%
\providecommand \enquote  [1]{``#1''}%
\providecommand \bibnamefont  [1]{#1}%
\providecommand \bibfnamefont [1]{#1}%
\providecommand \citenamefont [1]{#1}%
\providecommand \href@noop [0]{\@secondoftwo}%
\providecommand \href [0]{\begingroup \@sanitize@url \@href}%
\providecommand \@href[1]{\@@startlink{#1}\@@href}%
\providecommand \@@href[1]{\endgroup#1\@@endlink}%
\providecommand \@sanitize@url [0]{\catcode `\\12\catcode `\$12\catcode
  `\&12\catcode `\#12\catcode `\^12\catcode `\_12\catcode `\%12\relax}%
\providecommand \@@startlink[1]{}%
\providecommand \@@endlink[0]{}%
\providecommand \url  [0]{\begingroup\@sanitize@url \@url }%
\providecommand \@url [1]{\endgroup\@href {#1}{\urlprefix }}%
\providecommand \urlprefix  [0]{URL }%
\providecommand \Eprint [0]{\href }%
\providecommand \doibase [0]{http://dx.doi.org/}%
\providecommand \selectlanguage [0]{\@gobble}%
\providecommand \bibinfo  [0]{\@secondoftwo}%
\providecommand \bibfield  [0]{\@secondoftwo}%
\providecommand \translation [1]{[#1]}%
\providecommand \BibitemOpen [0]{}%
\providecommand \bibitemStop [0]{}%
\providecommand \bibitemNoStop [0]{.\EOS\space}%
\providecommand \EOS [0]{\spacefactor3000\relax}%
\providecommand \BibitemShut  [1]{\csname bibitem#1\endcsname}%
\let\auto@bib@innerbib\@empty
\bibitem [{\citenamefont {Will}(2014)}]{lrr-2014-4}%
  \BibitemOpen
  \bibfield  {author} {\bibinfo {author} {\bibfnamefont {C.~M.}\ \bibnamefont
  {Will}},\ }\href {\doibase 10.12942/lrr-2014-4} {\bibfield  {journal}
  {\bibinfo  {journal} {Living Reviews in Relativity}\ }\textbf {\bibinfo
  {volume} {17}} (\bibinfo {year} {2014}),\ 10.12942/lrr-2014-4},\ \Eprint
  {http://arxiv.org/abs/1403.7377} {arXiv:1403.7377 [gr-qc]} \BibitemShut
  {NoStop}%
\bibitem [{\citenamefont {Hawking}(1974)}]{Hawking:1974rv}%
  \BibitemOpen
  \bibfield  {author} {\bibinfo {author} {\bibfnamefont {S.}~\bibnamefont
  {Hawking}},\ }\href {\doibase 10.1038/248030a0} {\bibfield  {journal}
  {\bibinfo  {journal} {Nature}\ }\textbf {\bibinfo {volume} {248}},\ \bibinfo
  {pages} {30} (\bibinfo {year} {1974})}\BibitemShut {NoStop}%
\bibitem [{\citenamefont {Almheiri}\ \emph {et~al.}(2013)\citenamefont
  {Almheiri}, \citenamefont {Marolf}, \citenamefont {Polchinski},\ and\
  \citenamefont {Sully}}]{Almheiri:2012rt}%
  \BibitemOpen
  \bibfield  {author} {\bibinfo {author} {\bibfnamefont {A.}~\bibnamefont
  {Almheiri}}, \bibinfo {author} {\bibfnamefont {D.}~\bibnamefont {Marolf}},
  \bibinfo {author} {\bibfnamefont {J.}~\bibnamefont {Polchinski}}, \ and\
  \bibinfo {author} {\bibfnamefont {J.}~\bibnamefont {Sully}},\ }\href
  {\doibase 10.1007/JHEP02(2013)062} {\bibfield  {journal} {\bibinfo  {journal}
  {JHEP}\ }\textbf {\bibinfo {volume} {1302}},\ \bibinfo {pages} {062}
  (\bibinfo {year} {2013})},\ \Eprint {http://arxiv.org/abs/1207.3123}
  {arXiv:1207.3123 [hep-th]} \BibitemShut {NoStop}%
\bibitem [{\citenamefont {Braunstein}\ \emph {et~al.}(2013)\citenamefont
  {Braunstein}, \citenamefont {Pirandola},\ and\ \citenamefont
  {{\.Z}yczkowski}}]{Braunstein:2009my}%
  \BibitemOpen
  \bibfield  {author} {\bibinfo {author} {\bibfnamefont {S.~L.}\ \bibnamefont
  {Braunstein}}, \bibinfo {author} {\bibfnamefont {S.}~\bibnamefont
  {Pirandola}}, \ and\ \bibinfo {author} {\bibfnamefont {K.}~\bibnamefont
  {{\.Z}yczkowski}},\ }\href {\doibase 10.1103/PhysRevLett.110.101301}
  {\bibfield  {journal} {\bibinfo  {journal} {Phys.Rev.Lett.}\ }\textbf
  {\bibinfo {volume} {110}},\ \bibinfo {pages} {101301} (\bibinfo {year}
  {2013})},\ \Eprint {http://arxiv.org/abs/0907.1190} {arXiv:0907.1190
  [quant-ph]} \BibitemShut {NoStop}%
\bibitem [{\citenamefont {Jacobson}(1995)}]{Jacobson:1995ab}%
  \BibitemOpen
  \bibfield  {author} {\bibinfo {author} {\bibfnamefont {T.}~\bibnamefont
  {Jacobson}},\ }\href {\doibase 10.1103/PhysRevLett.75.1260} {\bibfield
  {journal} {\bibinfo  {journal} {Phys.Rev.Lett.}\ }\textbf {\bibinfo {volume}
  {75}},\ \bibinfo {pages} {1260} (\bibinfo {year} {1995})},\ \Eprint
  {http://arxiv.org/abs/gr-qc/9504004} {arXiv:gr-qc/9504004 [gr-qc]}
  \BibitemShut {NoStop}%
\bibitem [{\citenamefont {Damour}\ and\ \citenamefont
  {Esposito-Farese}(1992)}]{Damour:1992we}%
  \BibitemOpen
  \bibfield  {author} {\bibinfo {author} {\bibfnamefont {T.}~\bibnamefont
  {Damour}}\ and\ \bibinfo {author} {\bibfnamefont {G.}~\bibnamefont
  {Esposito-Farese}},\ }\href {\doibase 10.1088/0264-9381/9/9/015} {\bibfield
  {journal} {\bibinfo  {journal} {Class.Quant.Grav.}\ }\textbf {\bibinfo
  {volume} {9}},\ \bibinfo {pages} {2093} (\bibinfo {year} {1992})}\BibitemShut
  {NoStop}%
\bibitem [{\citenamefont {Kanti}\ \emph {et~al.}(1996)\citenamefont {Kanti},
  \citenamefont {Mavromatos}, \citenamefont {Rizos}, \citenamefont {Tamvakis},\
  and\ \citenamefont {Winstanley}}]{Kanti:1995vq}%
  \BibitemOpen
  \bibfield  {author} {\bibinfo {author} {\bibfnamefont {P.}~\bibnamefont
  {Kanti}}, \bibinfo {author} {\bibfnamefont {N.~E.}\ \bibnamefont
  {Mavromatos}}, \bibinfo {author} {\bibfnamefont {J.}~\bibnamefont {Rizos}},
  \bibinfo {author} {\bibfnamefont {K.}~\bibnamefont {Tamvakis}}, \ and\
  \bibinfo {author} {\bibfnamefont {E.}~\bibnamefont {Winstanley}},\ }\href
  {\doibase 10.1103/PhysRevD.54.5049} {\bibfield  {journal} {\bibinfo
  {journal} {Phys.Rev.}\ }\textbf {\bibinfo {volume} {D54}},\ \bibinfo {pages}
  {5049} (\bibinfo {year} {1996})},\ \Eprint
  {http://arxiv.org/abs/hep-th/9511071} {arXiv:hep-th/9511071 [hep-th]}
  \BibitemShut {NoStop}%
\bibitem [{\citenamefont {de~Rham}(2014)}]{deRham:2014zqa}%
  \BibitemOpen
  \bibfield  {author} {\bibinfo {author} {\bibfnamefont {C.}~\bibnamefont
  {de~Rham}},\ }\href@noop {} {\  (\bibinfo {year} {2014})},\ \Eprint
  {http://arxiv.org/abs/1401.4173} {arXiv:1401.4173 [hep-th]} \BibitemShut
  {NoStop}%
\bibitem [{\citenamefont {Jackiw}\ and\ \citenamefont
  {Pi}(2003)}]{Jackiw:2003pm}%
  \BibitemOpen
  \bibfield  {author} {\bibinfo {author} {\bibfnamefont {R.}~\bibnamefont
  {Jackiw}}\ and\ \bibinfo {author} {\bibfnamefont {S.~Y.}\ \bibnamefont
  {Pi}},\ }\href {\doibase 10.1103/PhysRevD.68.104012} {\bibfield  {journal}
  {\bibinfo  {journal} {Phys.Rev.}\ }\textbf {\bibinfo {volume} {D68}},\
  \bibinfo {pages} {104012} (\bibinfo {year} {2003})},\ \Eprint
  {http://arxiv.org/abs/gr-qc/0308071} {arXiv:gr-qc/0308071 [gr-qc]}
  \BibitemShut {NoStop}%
\bibitem [{\citenamefont {Alexander}\ and\ \citenamefont
  {Yunes}(2009)}]{Alexander:2009tp}%
  \BibitemOpen
  \bibfield  {author} {\bibinfo {author} {\bibfnamefont {S.}~\bibnamefont
  {Alexander}}\ and\ \bibinfo {author} {\bibfnamefont {N.}~\bibnamefont
  {Yunes}},\ }\href {\doibase 10.1016/j.physrep.2009.07.002} {\bibfield
  {journal} {\bibinfo  {journal} {Phys.Rept.}\ }\textbf {\bibinfo {volume}
  {480}},\ \bibinfo {pages} {1} (\bibinfo {year} {2009})},\ \Eprint
  {http://arxiv.org/abs/0907.2562} {arXiv:0907.2562 [hep-th]} \BibitemShut
  {NoStop}%
\bibitem [{\citenamefont {Delsate}\ \emph {et~al.}(2014)\citenamefont
  {Delsate}, \citenamefont {Hilditch},\ and\ \citenamefont
  {Witek}}]{Delsate:2014hba}%
  \BibitemOpen
  \bibfield  {author} {\bibinfo {author} {\bibfnamefont {T.}~\bibnamefont
  {Delsate}}, \bibinfo {author} {\bibfnamefont {D.}~\bibnamefont {Hilditch}}, \
  and\ \bibinfo {author} {\bibfnamefont {H.}~\bibnamefont {Witek}},\
  }\href@noop {} {\  (\bibinfo {year} {2014})},\ \Eprint
  {http://arxiv.org/abs/1407.6727} {arXiv:1407.6727 [gr-qc]} \BibitemShut
  {NoStop}%
\bibitem [{\citenamefont {Yunes}\ and\ \citenamefont
  {Pretorius}(2009)}]{Yunes:2009gi}%
  \BibitemOpen
  \bibfield  {author} {\bibinfo {author} {\bibfnamefont {N.}~\bibnamefont
  {Yunes}}\ and\ \bibinfo {author} {\bibfnamefont {F.}~\bibnamefont
  {Pretorius}},\ }\href {\doibase 10.1103/PhysRevD.79.084043} {\bibfield
  {journal} {\bibinfo  {journal} {Phys. Rev. D}\ }\textbf {\bibinfo {volume}
  {79}},\ \bibinfo {pages} {084043} (\bibinfo {year} {2009})},\ \Eprint
  {http://arxiv.org/abs/0902.4669} {arXiv:0902.4669 [gr-qc]} \BibitemShut
  {NoStop}%
\bibitem [{\citenamefont {Konno}\ \emph {et~al.}(2009)\citenamefont {Konno},
  \citenamefont {Matsuyama},\ and\ \citenamefont {Tanda}}]{Konno:2009kg}%
  \BibitemOpen
  \bibfield  {author} {\bibinfo {author} {\bibfnamefont {K.}~\bibnamefont
  {Konno}}, \bibinfo {author} {\bibfnamefont {T.}~\bibnamefont {Matsuyama}}, \
  and\ \bibinfo {author} {\bibfnamefont {S.}~\bibnamefont {Tanda}},\ }\href
  {\doibase 10.1143/PTP.122.561} {\bibfield  {journal} {\bibinfo  {journal}
  {Prog.Theor.Phys.}\ }\textbf {\bibinfo {volume} {122}},\ \bibinfo {pages}
  {561} (\bibinfo {year} {2009})},\ \Eprint {http://arxiv.org/abs/0902.4767}
  {arXiv:0902.4767 [gr-qc]} \BibitemShut {NoStop}%
\bibitem [{\citenamefont {Pani}\ \emph {et~al.}(2011)\citenamefont {Pani},
  \citenamefont {Macedo}, \citenamefont {Crispino},\ and\ \citenamefont
  {Cardoso}}]{Pani:2011gy}%
  \BibitemOpen
  \bibfield  {author} {\bibinfo {author} {\bibfnamefont {P.}~\bibnamefont
  {Pani}}, \bibinfo {author} {\bibfnamefont {C.~F.~B.}\ \bibnamefont {Macedo}},
  \bibinfo {author} {\bibfnamefont {L.~C.~B.}\ \bibnamefont {Crispino}}, \ and\
  \bibinfo {author} {\bibfnamefont {V.}~\bibnamefont {Cardoso}},\ }\href
  {\doibase 10.1103/PhysRevD.84.087501} {\bibfield  {journal} {\bibinfo
  {journal} {Phys.Rev.}\ }\textbf {\bibinfo {volume} {D84}},\ \bibinfo {pages}
  {087501} (\bibinfo {year} {2011})},\ \Eprint {http://arxiv.org/abs/1109.3996}
  {arXiv:1109.3996 [gr-qc]} \BibitemShut {NoStop}%
\bibitem [{\citenamefont {Yagi}\ \emph
  {et~al.}(2012{\natexlab{a}})\citenamefont {Yagi}, \citenamefont {Yunes},\
  and\ \citenamefont {Tanaka}}]{Yagi:2012ya}%
  \BibitemOpen
  \bibfield  {author} {\bibinfo {author} {\bibfnamefont {K.}~\bibnamefont
  {Yagi}}, \bibinfo {author} {\bibfnamefont {N.}~\bibnamefont {Yunes}}, \ and\
  \bibinfo {author} {\bibfnamefont {T.}~\bibnamefont {Tanaka}},\ }\href
  {\doibase 10.1103/PhysRevD.86.044037} {\bibfield  {journal} {\bibinfo
  {journal} {Phys.Rev.}\ }\textbf {\bibinfo {volume} {D86}},\ \bibinfo {pages}
  {044037} (\bibinfo {year} {2012}{\natexlab{a}})},\ \Eprint
  {http://arxiv.org/abs/1206.6130} {arXiv:1206.6130 [gr-qc]} \BibitemShut
  {NoStop}%
\bibitem [{\citenamefont {Konno}\ and\ \citenamefont
  {Takahashi}(2014)}]{Konno:2014qua}%
  \BibitemOpen
  \bibfield  {author} {\bibinfo {author} {\bibfnamefont {K.}~\bibnamefont
  {Konno}}\ and\ \bibinfo {author} {\bibfnamefont {R.}~\bibnamefont
  {Takahashi}},\ }\href@noop {} {\  (\bibinfo {year} {2014})},\ \Eprint
  {http://arxiv.org/abs/1406.0957} {arXiv:1406.0957 [gr-qc]} \BibitemShut
  {NoStop}%
\bibitem [{\citenamefont {Kleihaus}\ \emph {et~al.}(2011)\citenamefont
  {Kleihaus}, \citenamefont {Kunz},\ and\ \citenamefont
  {Radu}}]{Kleihaus:2011tg}%
  \BibitemOpen
  \bibfield  {author} {\bibinfo {author} {\bibfnamefont {B.}~\bibnamefont
  {Kleihaus}}, \bibinfo {author} {\bibfnamefont {J.}~\bibnamefont {Kunz}}, \
  and\ \bibinfo {author} {\bibfnamefont {E.}~\bibnamefont {Radu}},\ }\href
  {\doibase 10.1103/PhysRevLett.106.151104} {\bibfield  {journal} {\bibinfo
  {journal} {Phys.Rev.Lett.}\ }\textbf {\bibinfo {volume} {106}},\ \bibinfo
  {pages} {151104} (\bibinfo {year} {2011})},\ \Eprint
  {http://arxiv.org/abs/1101.2868} {arXiv:1101.2868 [gr-qc]} \BibitemShut
  {NoStop}%
\bibitem [{\citenamefont {{Wald}}(1984)}]{Wald}%
  \BibitemOpen
  \bibfield  {author} {\bibinfo {author} {\bibfnamefont {R.~M.}\ \bibnamefont
  {{Wald}}},\ }\href@noop {} {\emph {\bibinfo {title} {{General Relativity}}}}\
  (\bibinfo  {publisher} {University of Chicago Press},\ \bibinfo {year}
  {1984})\BibitemShut {NoStop}%
\bibitem [{\citenamefont {Stein}\ and\ \citenamefont
  {Yagi}(2014)}]{Stein:2013wza}%
  \BibitemOpen
  \bibfield  {author} {\bibinfo {author} {\bibfnamefont {L.~C.}\ \bibnamefont
  {Stein}}\ and\ \bibinfo {author} {\bibfnamefont {K.}~\bibnamefont {Yagi}},\
  }\href {\doibase 10.1103/PhysRevD.89.044026} {\bibfield  {journal} {\bibinfo
  {journal} {Phys.Rev.}\ }\textbf {\bibinfo {volume} {D89}},\ \bibinfo {pages}
  {044026} (\bibinfo {year} {2014})},\ \Eprint {http://arxiv.org/abs/1310.6743}
  {arXiv:1310.6743 [gr-qc]} \BibitemShut {NoStop}%
\bibitem [{\citenamefont {Campbell}\ \emph {et~al.}(1990)\citenamefont
  {Campbell}, \citenamefont {Duncan}, \citenamefont {Kaloper},\ and\
  \citenamefont {Olive}}]{Campbell:1990ai}%
  \BibitemOpen
  \bibfield  {author} {\bibinfo {author} {\bibfnamefont {B.~A.}\ \bibnamefont
  {Campbell}}, \bibinfo {author} {\bibfnamefont {M.~J.}\ \bibnamefont
  {Duncan}}, \bibinfo {author} {\bibfnamefont {N.}~\bibnamefont {Kaloper}}, \
  and\ \bibinfo {author} {\bibfnamefont {K.~A.}\ \bibnamefont {Olive}},\ }\href
  {\doibase 10.1016/0370-2693(90)90227-W} {\bibfield  {journal} {\bibinfo
  {journal} {Phys.Lett.}\ }\textbf {\bibinfo {volume} {B251}},\ \bibinfo
  {pages} {34} (\bibinfo {year} {1990})}\BibitemShut {NoStop}%
\bibitem [{\citenamefont {Teukolsky}(1972)}]{Teukolsky:1972my}%
  \BibitemOpen
  \bibfield  {author} {\bibinfo {author} {\bibfnamefont {S.}~\bibnamefont
  {Teukolsky}},\ }\href {\doibase 10.1103/PhysRevLett.29.1114} {\bibfield
  {journal} {\bibinfo  {journal} {Phys.Rev.Lett.}\ }\textbf {\bibinfo {volume}
  {29}},\ \bibinfo {pages} {1114} (\bibinfo {year} {1972})}\BibitemShut
  {NoStop}%
\bibitem [{\citenamefont {Teukolsky}(1973)}]{Teukolsky:1973ha}%
  \BibitemOpen
  \bibfield  {author} {\bibinfo {author} {\bibfnamefont {S.~A.}\ \bibnamefont
  {Teukolsky}},\ }\href {\doibase 10.1086/152444} {\bibfield  {journal}
  {\bibinfo  {journal} {Astrophys.J.}\ }\textbf {\bibinfo {volume} {185}},\
  \bibinfo {pages} {635} (\bibinfo {year} {1973})}\BibitemShut {NoStop}%
\bibitem [{\citenamefont {{Teukolsky}}(1974)}]{1974PhDT.......103T}%
  \BibitemOpen
  \bibfield  {author} {\bibinfo {author} {\bibfnamefont {S.~A.}\ \bibnamefont
  {{Teukolsky}}},\ }\emph {\bibinfo {title} {{Perturbations of a rotating black
  hole}}},\ \href {http://thesis.library.caltech.edu/2997/} {Ph.D. thesis},\
  \bibinfo  {school} {California Institute of Technology} (\bibinfo {year}
  {1974})\BibitemShut {NoStop}%
\bibitem [{\citenamefont {Stein}(2014)}]{Stein:2014wza}%
  \BibitemOpen
  \bibfield  {author} {\bibinfo {author} {\bibfnamefont {L.~C.}\ \bibnamefont
  {Stein}},\ }\href@noop {} {\  (\bibinfo {year} {2014})},\ \Eprint
  {http://arxiv.org/abs/1407.0744} {arXiv:1407.0744 [gr-qc]} \BibitemShut
  {NoStop}%
\bibitem [{\citenamefont {Ottewill}\ and\ \citenamefont
  {Taylor}(2012)}]{Ottewill:2012aj}%
  \BibitemOpen
  \bibfield  {author} {\bibinfo {author} {\bibfnamefont {A.~C.}\ \bibnamefont
  {Ottewill}}\ and\ \bibinfo {author} {\bibfnamefont {P.}~\bibnamefont
  {Taylor}},\ }\href {\doibase 10.1103/PhysRevD.86.024036} {\bibfield
  {journal} {\bibinfo  {journal} {Phys.Rev.}\ }\textbf {\bibinfo {volume}
  {D86}},\ \bibinfo {pages} {024036} (\bibinfo {year} {2012})},\ \Eprint
  {http://arxiv.org/abs/1205.5587} {arXiv:1205.5587 [gr-qc]} \BibitemShut
  {NoStop}%
\bibitem [{\citenamefont {Taylor}(2013)}]{TaylorsEmail}%
  \BibitemOpen
  \bibfield  {author} {\bibinfo {author} {\bibfnamefont {P.}~\bibnamefont
  {Taylor}},\ }\href@noop {} {}\bibinfo {howpublished} {personal communication}
  (\bibinfo {year} {2013})\BibitemShut {NoStop}%
\bibitem [{\citenamefont {Boyd}(2001)}]{boyd2001chebyshev}%
  \BibitemOpen
  \bibfield  {author} {\bibinfo {author} {\bibfnamefont {J.}~\bibnamefont
  {Boyd}},\ }\href {http://books.google.com/books?id=lEWnQWyzLQYC} {\emph
  {\bibinfo {title} {Chebyshev and Fourier Spectral Methods: Second Revised
  Edition}}},\ Dover Books on Mathematics\ (\bibinfo  {publisher} {Dover
  Publications},\ \bibinfo {year} {2001})\BibitemShut {NoStop}%
\bibitem [{\citenamefont {Shafee}\ \emph {et~al.}(2006)\citenamefont {Shafee},
  \citenamefont {McClintock}, \citenamefont {Narayan}, \citenamefont {Davis},
  \citenamefont {Li} \emph {et~al.}}]{Shafee:2005ef}%
  \BibitemOpen
  \bibfield  {author} {\bibinfo {author} {\bibfnamefont {R.}~\bibnamefont
  {Shafee}}, \bibinfo {author} {\bibfnamefont {J.~E.}\ \bibnamefont
  {McClintock}}, \bibinfo {author} {\bibfnamefont {R.}~\bibnamefont {Narayan}},
  \bibinfo {author} {\bibfnamefont {S.~W.}\ \bibnamefont {Davis}}, \bibinfo
  {author} {\bibfnamefont {L.-X.}\ \bibnamefont {Li}},  \emph {et~al.},\ }\href
  {\doibase 10.1086/498938} {\bibfield  {journal} {\bibinfo  {journal}
  {Astrophys.J.}\ }\textbf {\bibinfo {volume} {636}},\ \bibinfo {pages} {L113}
  (\bibinfo {year} {2006})},\ \Eprint {http://arxiv.org/abs/astro-ph/0508302}
  {arXiv:astro-ph/0508302 [astro-ph]} \BibitemShut {NoStop}%
\bibitem [{\citenamefont {Ali-Haimoud}\ and\ \citenamefont
  {Chen}(2011)}]{AliHaimoud:2011fw}%
  \BibitemOpen
  \bibfield  {author} {\bibinfo {author} {\bibfnamefont {Y.}~\bibnamefont
  {Ali-Haimoud}}\ and\ \bibinfo {author} {\bibfnamefont {Y.}~\bibnamefont
  {Chen}},\ }\href {\doibase 10.1103/PhysRevD.84.124033} {\bibfield  {journal}
  {\bibinfo  {journal} {Phys.Rev.}\ }\textbf {\bibinfo {volume} {D84}},\
  \bibinfo {pages} {124033} (\bibinfo {year} {2011})},\ \Eprint
  {http://arxiv.org/abs/1110.5329} {arXiv:1110.5329 [astro-ph.HE]} \BibitemShut
  {NoStop}%
\bibitem [{\citenamefont {Yagi}\ \emph
  {et~al.}(2012{\natexlab{b}})\citenamefont {Yagi}, \citenamefont {Yunes},\
  and\ \citenamefont {Tanaka}}]{Yagi:2012vf}%
  \BibitemOpen
  \bibfield  {author} {\bibinfo {author} {\bibfnamefont {K.}~\bibnamefont
  {Yagi}}, \bibinfo {author} {\bibfnamefont {N.}~\bibnamefont {Yunes}}, \ and\
  \bibinfo {author} {\bibfnamefont {T.}~\bibnamefont {Tanaka}},\ }\href
  {\doibase 10.1103/PhysRevLett.109.251105} {\bibfield  {journal} {\bibinfo
  {journal} {Phys.Rev.Lett.}\ }\textbf {\bibinfo {volume} {109}},\ \bibinfo
  {pages} {251105} (\bibinfo {year} {2012}{\natexlab{b}})},\ \Eprint
  {http://arxiv.org/abs/1208.5102} {arXiv:1208.5102 [gr-qc]} \BibitemShut
  {NoStop}%
\end{thebibliography}%

\end{document}